\documentclass[%
 reprint,
 amsmath,amssymb,
 aps,
]{revtex4-2}

\usepackage{graphicx}
\usepackage{dcolumn}
\usepackage{bm}
\usepackage{siunitx}

\begin{document}

\preprint{APS/123-QED}

\title{Nanodomain poling unlocking backward nonlinear light generation in thin film lithium niobate}

\author{Alessandra Sabatti}
\altaffiliation{These authors contributed equally to this work}
\email{asabatti@ethz.ch}

\author{Jost Kellner}%
\altaffiliation{These authors contributed equally to this work.}

\author{Robert J. Chapman}

\author{Rachel Grange}

\affiliation{%
 Optical Nanomaterial Group, Institute for Quantum Electronics, Department of Physics, ETH Zurich, CH-8093 Zurich, Switzerland}%

\date{\today}

\begin{abstract}
Nonlinear frequency conversion offers powerful capabilities for applications in telecommunications, signal processing, and computing. Thin-film lithium niobate (TFLN) has emerged as a promising integrated photonics platform due to its strong electro-optic effect and second-order nonlinearity, which can be exploited through periodic poling. However, conventional poling techniques in x-cut TFLN are limited to minimum period sizes on the order of microns, preventing the efficient generation of interactions involving counter-propagating waves. Here we report scalable periodic poling of x-cut TFLN with periods down to \SI{215}{nm} and realize devices for counter- and back-propagating phase matching. We estimate conversion efficiencies of \SI{1474}{\%/W/cm^2} and \SI{45}{\%/W/cm^2} respectively, and measuring sum frequency generation we confirm that the nonlinear generation takes place in the desired direction. We report spontaneous parametric down conversion for the counter-propagating and, for the first time, for a backward propagating device. This technological advance provides the control of domain geometry in TFLN with an unprecedented precision and leads into the generation of photon pairs with spatial and spectral properties tailored for quantum signal processing, quantum computing and metrology.
\end{abstract}

\maketitle


Nonlinear optics enables a large variety of light–matter interactions, with applications in ultrafast signal processing, frequency comb generation, and advanced imaging and sensing technologies \cite{wang2018modulator, FreqComb, nonlinearmicroscopy, wang2023image, MidIR_OPO}. 
Efficient frequency conversion for three-wave mixing processes, such as spontaneous parametric down-conversion (SPDC), is achieved in ferroelectric nonlinear crystals by implementing quasi-phase matching using periodic poling \cite{LoncarPPLN,LNOI_spdc,SPDC_brightness}.
One of the most promising platforms for efficient nonlinear optics applications is thin film lithium niobate (TFLN) thanks to its large $\chi^{(2)}$ nonlinearity and electro-optic effect, along with recent developments in fabrication techniques \cite{chen2022advances,Loncar_review,chen_adapted_2024,WavelengthAccurate,FabPaper}.
Periodic poling of lithium niobate in conventional applications employs extended domain periods, ranging from \SI{2}{} to tens of micrometers, to counteract material dispersion and provide the necessary momentum for down-converted signal and idler photons. This causes the three waves to propagate in the same direction along the crystal.
Quasi-phase matching can be engineered to control the propagation direction of the generated waves, unlocking nonlinear interactions with unique spectral and spatial properties \cite{liu_narrowband_2021,canalias_mirrorless_2007, d1997nonlinear, PulseShaping}.
The down-conversion of a pump photon into a signal travelling forward and an idler travelling backward is referred to as counter-propagating. The photons generated in this process exhibit a high heralded purity, which is crucial for quantum interference, thus for quantum information processing and computing \cite{Graffitti_purity, HeraldedPhotons, domain_engineering_quantum,CP_coherence}. The deterministic separation of signal and idler, moreover, is advantageous for photon routing with applications in quantum networks and communication protocols \cite{PhotonRoute, finco2024time}. However, the phase matching scheme enabling counter-propagating down-conversion requires a periodic poling that compensates for the full $\vec{k}$ vector of the pump (for degenerate signal and idler). The resulting poling period is much shorter than the standard forward propagating period and is expressed by: $\Lambda = \lambda_p/n(\lambda_p)$, where $\lambda_p$ is the pump wavelength and $n(\lambda_p)$ is its effective refractive index. 
For a SPDC process generating photon pairs in the telecom C-band, in lithium niobate and for most photonics platforms, the period approaches \SI{400}{nm}, whose manufacturing is very challenging due to the lateral merging of the domains \cite{UV_chirp, UV_marandi, Nagy_submicron_poling}.
Another process enabled by an even shorter poling period involves the pump photon down-converting into two photons travelling backwards, that we will reference in the following as backward-propagating down-conversion. The generated photon pairs are extremely broadband, enabling precise time resolution for ultrafast quantum optics and time-bin entanglement, yielding frequency entangled photon pairs that are naturally filtered from the pump \cite{TimeBin, finco2024time, LNOI_spdc}. 
The poling period $\Lambda$ fulfils the following expression for degenerate signal and idler: $\Lambda = \lambda_p/[n(\lambda_p)+n(\lambda_s)]$ and requires around a hundred nanometre-sized domains. 

However, since these ultra-short periods set a huge challenge in fabrication, demonstrations in integrated photonic platforms, that benefit from larger nonlinear efficiencies with respect to bulk optics due to a reduced mode volume and stronger confinement, are still lacking. 
Counter-propagating SPDC was achieved in bulk periodically poled nonlinear crystals \cite{counterprop_PPKTP_spdc}, and in integrated platforms exploiting higher order poling to relax the constraint on the periodicity and consequently with reduced efficiency \cite{kellner2025counter,luo_counter-propagating_2020,liu_narrowband_2021,liu2021observation}. Realizations with first order periodic poling in an integrated platform are still missing and backward-propagating SPDC has not been experimentally demonstrated.

\begin{figure*}
    \centering
    \includegraphics[width=0.9\linewidth]{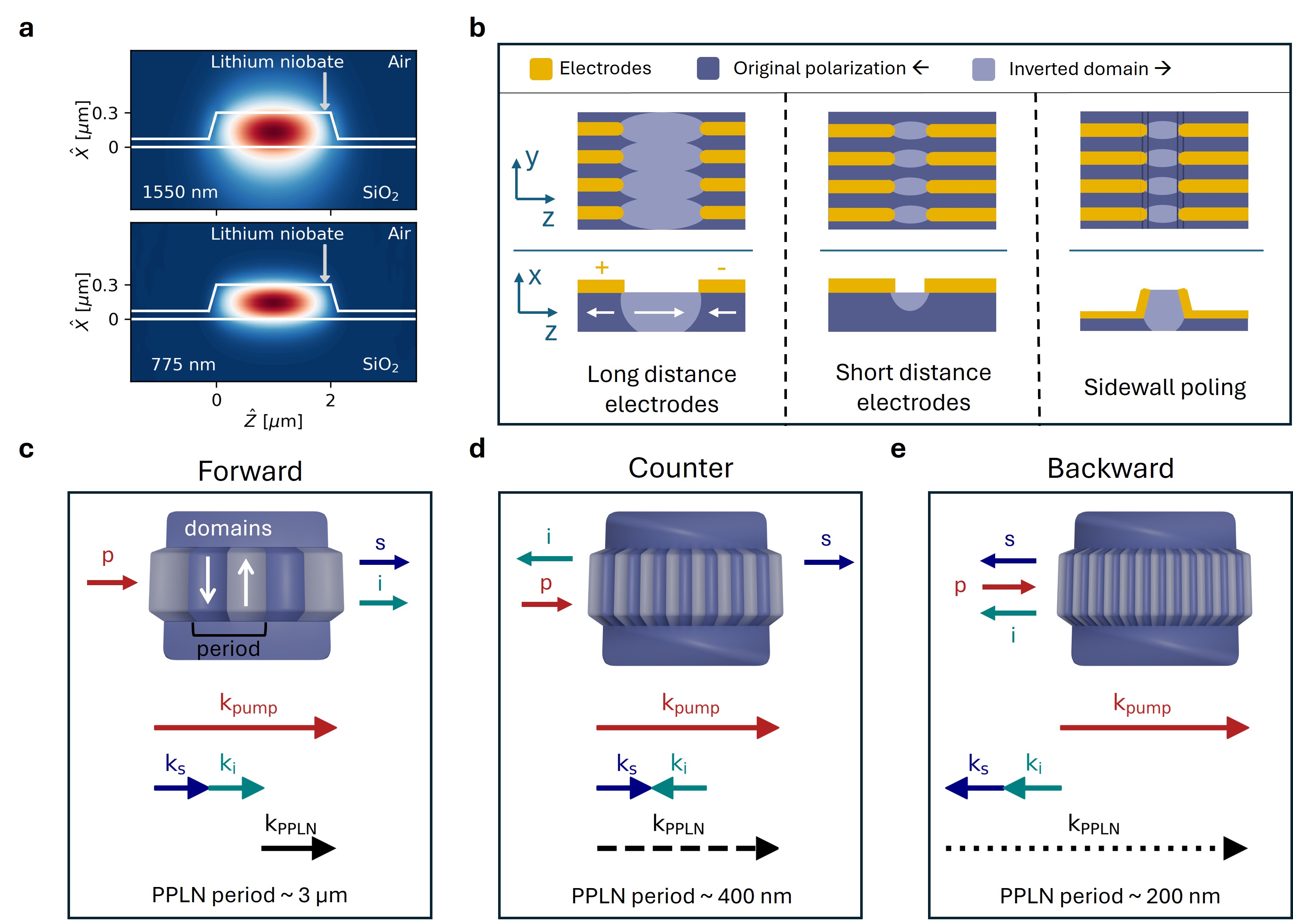}
    \caption{\textbf{Overview of periodic poling for phase matching processes requiring short periodicity}. \textbf{a} Waveguide geometry and optical modes involved in the phase matching in the thin film lithium niobate platform for \SI{1550}{nm} and \SI{775}{nm}. \textbf{b} Graphical representation of the periodic poling process for standard poling before etching and sidewall poling applied to ultrashort periods. With long distance electrode the domains merge laterally, whereas with short distance electrodes they do not reach the whole depth. The proposed solution using sidewall poling allows for limiting the lateral growth while obtaining a fully poled film. Spontaneous parametric down-conversion process represented in a poled waveguide and momentum mismatch for \textbf{c} standard all-forward propagating, \textbf{d} counter-propagating and \textbf{e} backward propagating signal and idler. White arrows on the waveguide indicate different polarization.   Abbreviations: p: pump, s: signal, i: idler, PPLN : periodic poling of lithium niobate.}
    \label{fig:fig1}
\end{figure*}

Here we realize scalable periodic poling of x-cut TFLN with periods of \SI{390}{nm} and \SI{215}{nm}. The domain growth is engineered by patterning the electrodes for high voltage application directly on the sidewalls of the waveguide \cite{franken_milliwatt-level_2025}. By measuring second harmonic generation, we estimate a high conversion efficiency of \SI{1470}{\%/W/cm^2} for the counter-propagating and we show excellent domain quality. For the backward wave device, we successfully measure backward second harmonic generation, reporting an efficiency of $45 \%/W/cm^2$. We perform sum frequency generation measurements and observe the characteristic phase matching functions, confirming that we are observing the desired phenomena. Lastly, we measure counter-propagating and, notably for the first time, backward propagating SPDC. This demonstration of ultrashort poling period shows that domain shape engineering can be performed with a scalable method at an unprecedented level of precision. Our work paves the way for new efficient approaches to integrated quantum information processing, quantum computing and metrology.


\section*{Phase matching with ultrashort domains}

The process of periodic poling with ultrashort period is investigated in an x-cut thin film lithium niobate (TFLN) sample with \SI{300}{nm} thick lithium niobate film on a silicon dioxide insulation layer. In the following we focus on implementing a fabrication technique that enables efficient counter-propagating and back-propagating phase matching. 
With the waveguide geometry represented in Fig.\ref{fig:fig1}a, and considering a pump photon at \SI{775}{nm}, the required poling periods for counter- and backward-propagating phase matching are $\sim$\SI{390}{nm} and $\sim$\SI{215}{nm} respectively. 
Achieving such short periods in x-cut TFLN is very challenging as the neighbouring domains tend to merge during the poling process, rendering phase matching unfeasible. 
The most widespread approach for achieving periodically poled waveguides is given by first patterning comb-like poling electrodes on the lithium niobate surface, next poling the film and last fabricating the waveguides in the poled region. This method is very successful for poling periods on the order of \SI{3}{\um} for traditional forward phase matching \cite{WavelengthAccurate}, but suffers from lateral domain merging at smaller periods \cite{rosenman1998domain, UV_chirp}. A highly efficient periodic poling must exhibit domain inversion in the whole film depth, extending in the x direction, and a 50\% duty cycle in the y direction, i.e. the light propagation direction. The periodic domains are generated in a nucleation process with a growth rate that is in general different in the z direction than in the x and y (ordinary axis) direction \cite{bulk_temperature}. As represented in Fig.\ref{fig:fig1}b, for sub-micron periods, if the two electrode combs are placed far apart, at tens of microns distance, as soon as the domain has fully grown in the z-direction, it has also spread in the whole film depth, but the lateral growth is excessive and results in merged domains \cite{rosenman1998domain}. If the electrodes are closer together, the lateral growth can be well calibrated, but it is not possible to reach the full film depth, resulting in low efficiency frequency conversion \cite{ruesing2019second}. To solve this problem, we use the technique of sidewall poling, presented by Franken et al. \cite{franken_milliwatt-level_2025}. Although it was realized to improve the previous \textit{poling-after-etching} techniques and enhance the conversion efficiency, we propose that it can be used to obtain domains widths on the order of a hundred nanometres. This method consists of etching the waveguide as a first step, and patterning the electrodes for poling directly on the waveguide sidewalls. Such a configuration allows nucleation to initiate not only on the film surface but also along the entire etched depth.

Fig.\ref{fig:fig1}c-e illustrate the influence of wave directionality in three phase matching processes and relate it to the required poling period in TFLN. The process of spontaneous parametric down-conversion (SPDC) is represented in terms of the propagation direction of the individual waves and the momentum mismatch, which is inversely proportional to the poling period.
The most well-known configuration involves all three waves propagating together, where a pump photon down-converts generating a signal and an idler at around half its frequency, both in the forward direction (Fig.\ref{fig:fig1}c). The momentum mismatch is only given by dispersion and it converts into a relatively large poling period of \SI{2.8}{\um}, which can be realized with the standard poling approach.
Fig.\ref{fig:fig1}d illustrates the counter-propagating scheme, with the pump converting into a forward-propagating signal and a backward-propagating idler. Near degeneracy for signal and idler frequencies, their momenta nearly cancel, and the mismatch corresponds to the momentum $\vec{\text{k}}$\textsubscript{sum} of the pump photon. This corresponds to a poling period of approximately half the pump wavelength. 
The last case under investigation is shown in Fig.\ref{fig:fig1}e, where the pump generates two backward-propagating photons. Here, the poling vector must compensate for the net momentum of all three waves. This requires an ultrashort poling period of \SI{215}{nm}, with domain widths as small as \SI{100}{nm}.

\section*{Fabrication and characterization}

\begin{figure*}
    \centering
    \includegraphics[width=1\linewidth]{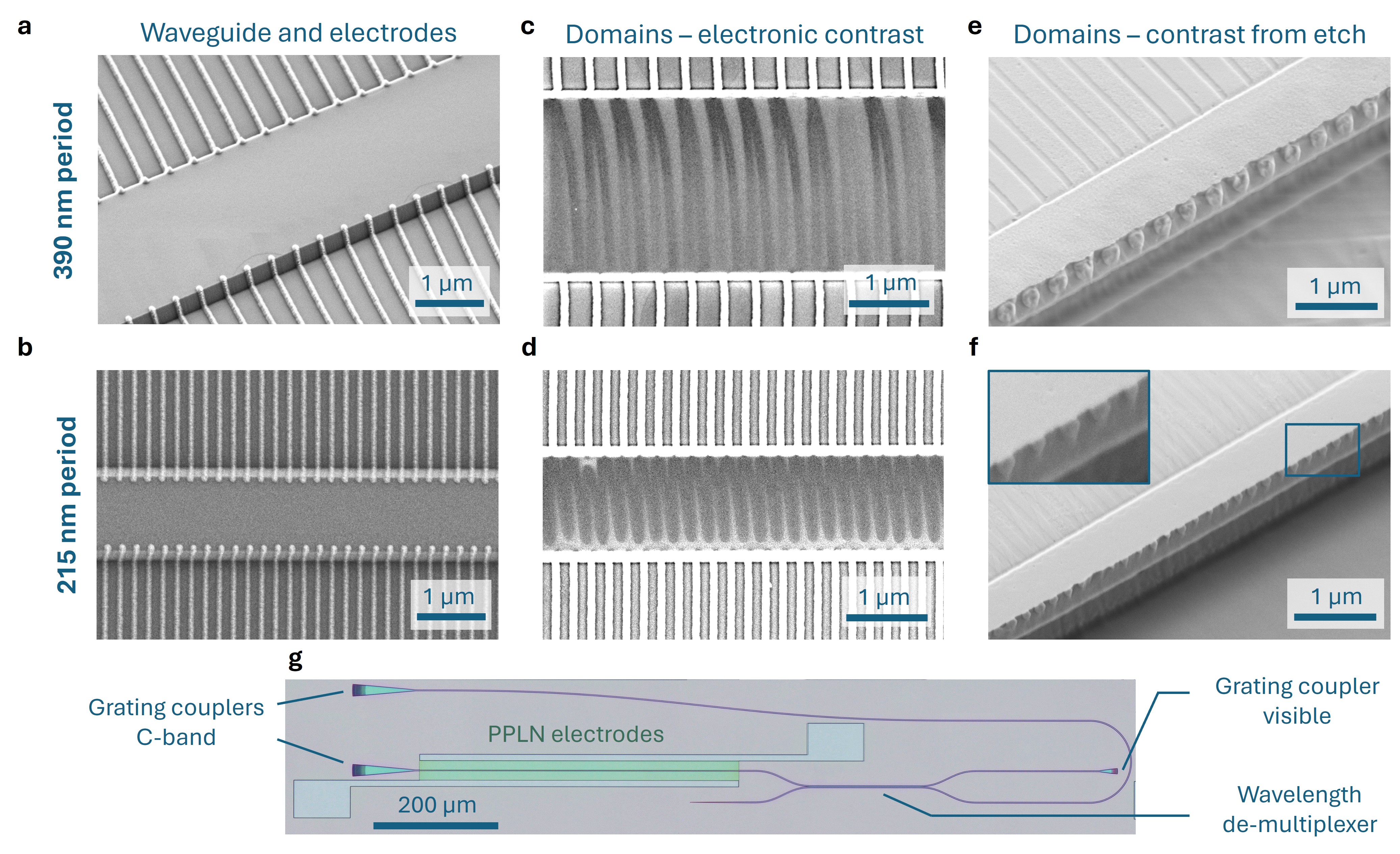}
    \caption{\textbf{Fabrication and characterization of ultrashort period poled waveguides}. \textbf{a} Waveguide with poling electrodes with \SI{390}{nm} period and \textbf{b} \si{215}{nm} period. SEM image of \textbf{c} \SI{390}{nm} and \textbf{d} \SI{215}{nm} period domains after poling. Wet etched waveguide with topological contrast between the domains for \textbf{e} \SI{390}{nm} and \textbf{f} \SI{215}{nm} periodic poling with inset showing zoomed in picture of the domains. \textbf{g} Optical microscope picture of a device including the periodically poled waveguide, a wavelength demultiplexer and grating couplers for inputs at \SI{1550}{nm} on the left and output at \SI{775}{nm} on the right.}
    \label{fig:fig2}
\end{figure*}

The sample is fabricated by etching the lithium niobate waveguides first and then depositing titanium electrodes directly on the structured film (details in Methods), in a \SI{0.5}{mm} long periodically poled lithium niobate (PPLN) region. The waveguide used for counter-propagating phase matching has a top width of \SI{2}{\um}, whereas the best result for backward-propagating phase matching was obtained with a reduced top width of \SI{1.15}{\um}. Fig\ref{fig:fig2}a-b show scanning electron microscope (SEM) images of the structures for \SI{390}{nm} and \SI{215}{nm} periods, respectively. The periodic poling is then performed via the application of one electric field pulse applied for \SI{1}{ms} with a peak voltage of \SI{80}{V} and \SI{130}{V} for the counter- and back-propagating devices, respectively. The outcome of the poling process is verified via SEM inspection. By tuning the acceleration voltage and beam current, it is possible to image ferroelectric domains, either obtaining a contrast between domains with different polarization, or an enhanced or suppressed signal at the domain wall, depending on the sample and imaging conditions \cite{SEMreview,Aristov1984}. In the images reported in Fig.\ref{fig:fig2}c-d, we obtain a signal contrast between the poled domains and the domain walls, such that the domain walls are darker than the domain volume.
For \SI{390}{nm} poling, the domains are well separated with a duty cycle close to 50 $\%$ (Fig.\ref{fig:fig2}c). Note that the curved shape of the domains close to the top of the waveguide is due to electron beam drifting induced by sample charging during image acquisition. Fig.\ref{fig:fig2}d, for a period of \SI{215}{nm}, shows the domain inversion with well-defined boundaries, however, the domains are merged at the bottom. 
The SEM images are acquired using secondary electrons with an acceleration voltage of \SI{2}{keV}, therefore the detected signal is expected to be emitted only from a volume a few nanometres into the surface, thus revealing little information about the poling depth.
To investigate the geometry of the domains in the full depth of the lithium niobate film, we etch the waveguide parallel to the edges and perform a wet etching step in a chemical that etches the surface at a different rate depending on the z-orientation of the crystal (details in Methods). Fig.\ref{fig:fig2}e shows clear well separated domains for PPLN with \SI{390}{nm} period, where the only deviation from ideality is the rounded cross-section instead of a square one. For the sample with \SI{215}{nm} period the domains have a duty cycle close to 50\% on the surface, but reach only about half of the film thickness (Fig.\ref{fig:fig2}e). This can be explained by the reduced waveguide top width adopted for the smaller periods. With the \SI{2}{\um} wide waveguide SEM inspection shows that it is not possible to achieve well separated domains, indicating that the lateral growth is too fast. To obtain a reduced lateral growth, we diminish the electrode distance and the top width accordingly. With closer electrodes, however, the inversion does not involve the entire film depth. 
A possible solution to achieve deeper domains is, after performing poling with this method, etching the portion of the electrodes lying on the sidewall, and performing a second poling to grow the domain in the remaining film.
In Fig.\ref{fig:fig2}g, an optical microscope picture of the fabricated waveguide and electrodes is displayed to show the circuit structure. Connected to the periodically poled waveguide, there is a wavelength demultiplexer, used to route pump, signal and idler to their given in and outputs. The demultiplexer has the structure of a directional coupler that fully couples light at the signal and idler wavelength from one waveguide to the other, while keeping the light at \SI{775}{nm} in the same waveguide. On the left the grating couplers serve as an input for the signal and idler at \SI{1550}{nm} and the output for the generated \SI{775}{nm} light is on the right.



\section*{Nonlinear spectrum and efficiency measurement}

\begin{figure*}
    \centering
    \includegraphics[width=1\linewidth]{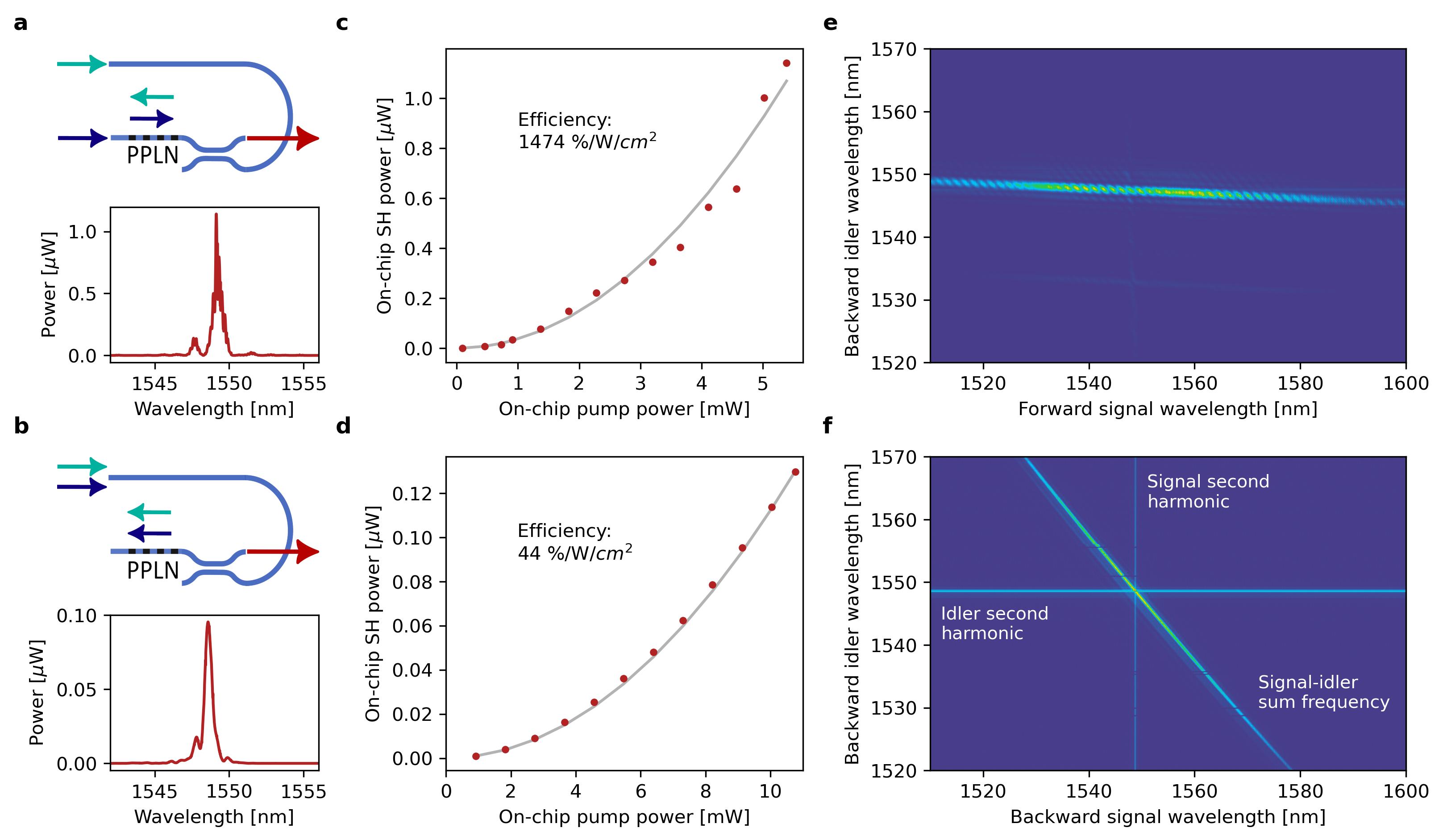}
    \caption{\textbf{Second harmonic and sum frequency measurements for counter-propagating and back-propagating signal and idler}. Circuit schematic and second harmonic spectrum for \textbf{a} counter- and \textbf{b} back-propagating. Second harmonic conversion efficiency for \textbf{c} counter- and \textbf{d} back-propagating calculated from input power sweep and quadratic fit. Experimental sum frequency generation maps for \textbf{e} counter- and \textbf{f} back-propagating device. The diagonal line represents the sum frequency coming form signal and idler, whereas the horizontal and vertical lines are the second harmonic produced by the idler and signal independently.}
    \label{fig:fig3}
\end{figure*}

The nonlinear optical response of the devices is characterized by first measuring the second harmonic signal, which allows us to estimate the conversion efficiency. We also perform a sum frequency scan as a function of the signal and idler frequencies to map the efficiency of the frequency conversion. This data is referred to as phase matching function and it has a characteristic shape for each of the three phase matching processes that we illustrated in Fig.\ref{fig:fig1} \cite{luo_counter-propagating_2020}. With these measurements it is further proved that the intended counter-propagating and backward-propagating phenomena are observed, and excluded that the backward propagation is generated by reflections in the circuit.

The second harmonic measurement is carried out using a continuous wave tunable laser centred around \SI{1550}{nm}. For the counter-propagating device, the light is split into two fibres that are then input to the chip, as represented by the circuit schematics in Fig.\ref{fig:fig3}a. The top waveguide is routed such that the signal propagates in the opposite direction with respect to the idler. The second harmonic signal is collected from the output on the right and its power displays a peak close to a pump value of \SI{1550}{nm} (Fig.\ref{fig:fig3}a). The efficiency is estimated by repeating this measurement for different pump powers. The second harmonic power follows a quadratic trend with respect to the pump power, as shown in Fig.\ref{fig:fig3}c, and the conversion efficiency estimated from the fit is \SI{1474}{\%/W/cm^2}. The value is about half of the theoretical efficiency of \SI{3020}{\%/W/cm^2}, corroborating the quality of the periodic poling. 
For the device with backward-propagating signal and idler, we followed a similar procedure, except that the laser power was entirely input into the top waveguide. This allows us to collect at the output the second-harmonic power generated in the opposite direction (Fig\ref{fig:fig3}b). The efficiency resulting from the power sweep and quadratic fit in Fig.\ref{fig:fig3}d is \SI{44}{\%/W/cm^2}. This reduction with respect to the theoretical value, we attribute to the non-ideal shape of the domains, in particular to their partial extension into the film in the x-direction. Note that the theoretical conversion efficiency is the same for all three phase matching schemes.

The sum frequency functions are acquired by sweeping the wavelength of two different lasers. The counter-propagating map displayed in Fig.\ref{fig:fig3}e shows a weak dependence of the forward signal on the wavelength. On the contrary, for the backward idler efficient phase matching takes place in a very narrow spectral range. This behaviour matches precisely the simulation (see Supplementary) and is peculiar of counter-propagating phase matching. This feature in an SPDC process enables near spectrally uncorrelated photon pairs for high purity heralded photon generation \cite{kellner2025counter}. For the back-propagating case, the data show three different lines (Fig.\ref{fig:fig3}f). The measured phase matching function is the diagonal line and presents a negative slope and narrowband spectrum, as expected from simulation (see Supplementary). While the diagonal line corresponds to the sum frequency involving photons from the two different lasers, the horizontal and vertical lines are the second harmonic signals produced by photons from the same laser. Since this type of phase matching involves two signal and idler photons co-propagating with respect to each other, when phase-matched, two photons from the same laser can generate a second harmonic photon. As expected, this is not observed in the case of counter-propagating pumps, except for very weak horizontal and vertical lines due to reflections of the signal and idler at the opposite inputs. 
We also note that for the back-propagating sum frequency, the map is invariant under the exchange of its variables, signal and idler, that both propagate backwards with respect to the sum. On the contrary, for counter-propagating phase matching, the forward- and backward-propagating signal and idler play different roles, as the sum frequency signal propagates forward (the signal direction) and breaks the symmetry.
The detailed measurement setup is described in the Methods section. 

\section*{Spontaneous parametric down-conversion}
\begin{figure}
    \centering
    \includegraphics[width=1\linewidth]{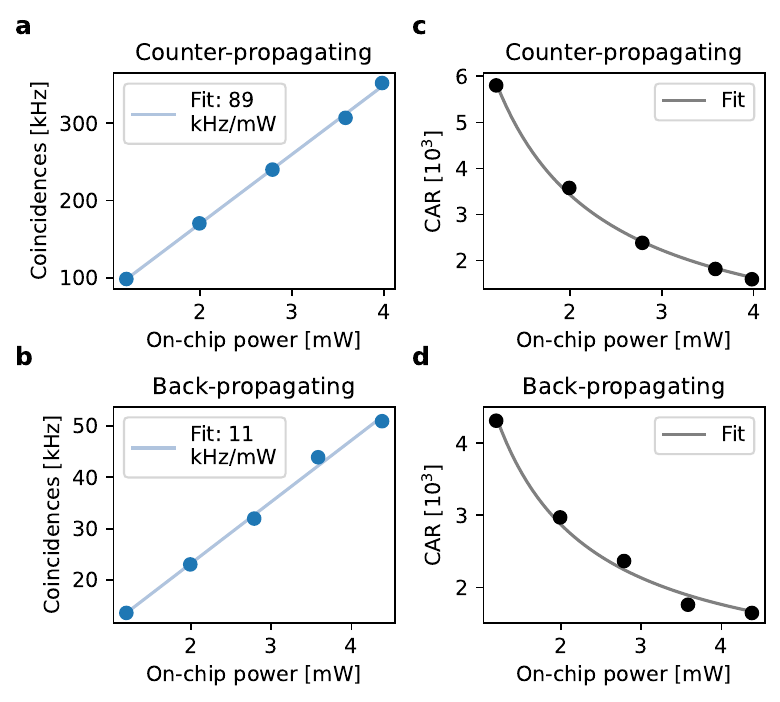}
    \caption{\textbf{Spontaneous parametric down-conversion measurements.} \textbf{a-b} On-chip coincidence rate as a function of pump power on-chip with linear fit for \textbf{a} counter- and \textbf{b} backward-propagating photon pairs. \textbf{c-d} Coincidence to accidental ratio with fit proportional to the reciprocal of the input power for \textbf{c} counter-propagating and \textbf{d} backward-propagating sources.}
    \label{fig:fig4}
\end{figure}

Lastly we measure photon pairs produced by both the counter- and back-propagating devices. We input a continuous wave laser at about \SI{774}{nm} and measure the generated photons with superconducting nanowires single photon detectors. The coincidence events are recorded using a time tagger. From the coincidence count rate as a function of the on-chip pump power reported in Fig.\ref{fig:fig4}a, we estimate an internal brightness of the counter-propagating source of \SI{89}{kHz/mW}. Despite the fact that the PPLN length is only \SI{0.5}{mm}, the brightness is roughly twice then reported for a \SI{6}{mm} long bulk periodically poled KTP and four orders of magnitude higher than titanium indiffused lithium niobate waveguides with fifth order poling period \cite{counterprop_PPKTP_spdc,luo_counter-propagating_2020}. We measure SPDC also from the back-propagating source and report a brightness of \SI{11}{kHz/mW}. The coincidence to accidental ratio (CAR) for both experiments is reported in Fig\ref{fig:fig4}c-d with values around 3000 at \SI{2}{mW}. The CAR follows the reciprocal of the input pump power, as shown by the fitting function.

\section*{Conclusion}

In summary, we have presented a scalable periodic poling technique, which enables high quality poling of domains \SI{200}{nm} wide, with full inversion in the entire thin film depth and \SI{100}{nm} domains with partial inversion. In both cases we successfully measured the phase matching functions for sum frequency, confirming the presence of the desired phase matching configurations, matching accurately with simulations. The measured efficiencies are \SI{1474}{\%/W/cm^2} and \SI{44}{\%/W/cm^2} respectively for counter- and backward-propagating periodic poling. With respect to the counter-propagating sum frequency with first order poling reported in z-cut lithium niobate by Yang et al. \cite{yang_symmetric_2024}, we report a similar efficiency, with the advantage of the potential integration of this method with the most established x-cut platform, which allows the integration of high-speed electro-optic functions. Backward second harmonic generation has been reported by Yakar et al. exploiting all optical poling in silicon nitride with an efficiency three orders of magnitude lower than presented in this work and with non-permanent domain switching \cite{backback_Bres}. 
    
We have shown spontaneous parametric down-conversion for counter-propagating photons with a source brightness of \SI{89}{kHz/mW} and back-propagating with \SI{11}{kHz/mW}. This demonstration of ultrashort poling period shows that domain shape engineering can be performed with hundred nanometres precision. Unlike for methods such as femtosecond laser poling \cite{xu2022femtosecond}, this technique is scalable and can be applied to millimetres long devices at once. We demonstrate that on-chip photon sources can be engineered to obtain a new degree of freedom of propagation direction. This result is important as it intrinsically allows either splitting of signal and idler or filtering the pump from the generated photons, functions which are hindered for the forward-propagating photon pair generation method. This method for efficiently generating counter-propagating photon pairs not only allows the generation of high brightness and high purity photon pair sources for quantum information processing, but also opens the possibility of realizing  mirrorless and degeneracy-locked optical parametric oscillator in the x-cut thin film lithium niobate platform \cite{canalias_mirrorless_2007,yang2025degeneracy}.

\section*{Methods}

\textbf{Device fabrication.} 
The sample is realized using the TFLN platform, with a \SI{5}{\%} MgO doped LN film with \SI{300}{nm} thickness, and with a \SI{2}{\um} thick buried oxide layer.
The waveguides are patterned into an HSQ-16 mask using electron beam lithography (EBL). A \SI{200}{nm} etching is performed via inductively coupled plasma reactive ion etching (ICP-RIE) argon milling, and the redeposition is removed with a KOH wet etching process. The remaining mask is etched with a buffered HF dip. The sample is annealed in air at \SI{500}{\degree C} for two hours. The poling electrodes are defined with another EBL step, followed by \SI{100}{nm} titanium evaporation and liftoff. The poling is performed applying high voltage pulses, using a photoresist as an insulation layer to avoid air breakdown. After poling, the electrodes are removed using HF. We report that on a preliminary sample, after poling, another annealing step was performed and caused the complete deterioration of the measured nonlinear signal. Subsequent SEM inspection showed the absence of domains on previously working devices. This behaviour can be attributed to the back-switching of the domains favoured by the high temperature and could be exploited to obtain reconfigurable poled devices. 
For characterization of the inner domain structure, a layer of \SI{50}{nm} of glass is deposited on the chip to protect the waveguides from the subsequent wet etching. A photoresist is patterned via direct writing laser lithography and rectangular holes are defined parallel to the waveguide. The lithium niobate layer is fully etched as well as a small layer of the bottom SiO$_2$, in order to highlight the boundary between the two materials. The photoresist is removed and the sample is immersed into an RCA-SC1 solution for 20 minutes, to obtain a topographic contrast between the domains. Finally the top glass protection layer is removed using buffered HF.  


\textbf{Experimental setup.}
We employ two distinct setups to characterize the optical properties of the two sources. Sum frequency generation and second harmonic generation are measured using either two tunable lasers centred around \SI{1550}{nm} (EXFO T200s and Keysight N7776C), or a single laser, depending on the configuration. Light is coupled into the chip via fibre arrays, that allow to access two neighbouring grating couplers. The generated second harmonic light is separated from residual signal and idler light off-chip using optical filters and detected with a silicon photodetector. By fixing the wavelength of one laser and sweeping the other, we map the spectral phase matching response of the device.

For the SPDC measurement, a laser (Newport Velocity TLB-6712) at \SI{775}{nm} is coupled into the chip through a polarization controller. At the output, the photon pairs are collected using fibre arrays. Two long-pass filters are deployed to suppress residual pump light before guiding the photons to superconducting nanowire single-photon detectors (Single Quantum Eos). Coincidence statistics are obtained using a TimeTagger (Swabian Ultra).

\section*{Author contributions}
A.S and J.K contributed equally to this work. A.S, J.K. and R.C. conceived the idea. A.S designed, fabricated and characterized the sample. J.K. performed optical measurements. The manuscript was written by A.S and J.K with inputs from all authors. R.G supervised the project.

\section*{Competing interest}
The authors declare no competing financial or non financial interests.

\section*{Funding}
This work was supported by the Swiss National Science Foundation Grant (project number 206008).

\begin{acknowledgments}
We acknowledge support for the fabrication and the characterization of our samples from the Scientific Center of Optical and Electron Microscopy ScopeM and from the cleanroom facilities BRNC and FIRST of ETH Zurich and IBM Ruschlikon. We thank Dr. Antonis Olziersky for the support in the process of electron-beam lithography, and Giovanni Finco and Dr. Tummas Napoleon Arge for helping to revise the manuscript.
\end{acknowledgments}

\section*{Data availability statement}
Data supporting the findings of this study are available within the article and Supplementary Material. Raw data and analysis code are available from the corresponding author upon reasonable request.\\


\bibliography{CounterProp}

\newpage

\onecolumngrid



\end{document}